\documentstyle[12pt,aasms4]{article}
\def\day{{}$^{\rm d}$\llap{.}}
\def\msun{{M_\odot}}
\def\lsun{{L_\odot}}

\begin{document}

\title{THE GLOBULAR CLUSTER $\omega$ CENTAURI AND
THE OOSTERHOFF DICHOTOMY}

\author{Christine M. Clement and Jason Rowe}

\affil{Department of Astronomy, University of Toronto \\
Toronto, Ontario, M5S 3H8, CANADA\\
electronic mail: cclement@astro.utoronto.ca,rowe@astro.utoronto.ca}

\begin{abstract}
CCD observations obtained by the OGLE team for
128 RR Lyrae variables in $\omega$ Centauri have been analysed. The 
period-luminosity and period-amplitude plots indicate that, in addition to
fundamental (RRab) and first overtone (RRc)
pulsators, the $\omega$ Centauri RR Lyrae population seems to
include second overtone (RRe) and possibly third overtone pulsators.
The mean period for the 59 RRab stars is 0\day 649, for the 48 RRc stars, it is
0\day 383 and for the 21 RRe stars, it is 0\day 304. 
The mean periods
derived for the RRab and RRc stars are typical values for an Oosterhoff type
II (OoII) cluster. Nevertheless, the period amplitude plot also shows
that some of the RR Lyrae variables have `Oosterhoff type I' (OoI)
characteristics.
Most of the second overtone variables exhibit non-radial pulsations similar to
those recently detected in some of the RR Lyrae variables in the clusters
M55 and M5, in the galactic bulge and in the LMC.
Relative luminosities derived for the RRc variables
from Fourier coefficients  correlate with the observed apparent magnitudes. 
Masses for the RRc stars have been calculated from Fourier coefficients. 
A comparison of the derived masses for RRc stars in the four OoII clusters
$\omega$ Cen, M15, M55 and M68
indicates that the masses of the
RRc stars in M15 and M68 are almost $0.2\msun$ greater
than those in the other two.
Since M15 and M68 have a high frequency of RRd stars among their first
overtone pulsators, while none have been identified in $\omega$ Cen or M55.
this suggests that the double-mode pulsation phenomenon may be associated with
mass. 
Among the RRc variables in $\omega$ Cen, the OoII variables 
have lower derived masses and higher
luminosities than the OoI variables.
An application of the period-density
law to pairs of OoI and OoII RRab stars selected according to their position
in the period-amplitude plot also indicates that the OoII variables in general
have lower masses and higher luminosities.
These findings support the hypothesis that the RR Lyrae variables in OoII
systems are evolved HB stars that
spend their ZAHB phase on the blue side of the instability strip.

\end{abstract} 
%
%
\section{INTRODUCTION}

More than sixty years ago, Oosterhoff (1939) discovered that Galactic 
globular clusters
could be classified into two groups according to the frequency distribution
of the periods of their RR Lyrae variables. He suggested that the 
absolute median magnitude of the variables
may differ from one cluster to another, but since his result was based 
on data 
for only five clusters, an extension of the investigation
would be desirable. 
Subsequent studies by Oosterhoff (1944) and  van Agt \& Oosterhoff (1959)
showed that the division into two groups was indeed more general.\footnote
{Based on the data for 17 clusters, van Agt and Oosterhoff found 
that for
group I (OoI) clusters, the mean period of the
RRc variables $<P_c>$ is 0\day 319, the mean period of the
RRab variables $<P_{ab}>$ is 0\day 549, and the proportion of
the variables classified as type c  ($n_c/N$) is $0.17$.  The 
comparable figures for the 
OoII clusters are $<P_c>=0.371$, $<P_{ab}>=0.647$ and $n_c/N=0.47$.}
At the same time, Sandage (1958) demonstrated
that  a difference of $\Delta M_V=0.2$ between the
RR  Lyrae variables in the two groups could account for the observed
difference in the periods. 
Meanwhile, 
evidence for a connection between Oosterhoff type and heavy element abundance
was mounting. 
Arp (1955) reported that the spectra of giant stars in the
OoII clusters M15 and M92 had excessively weak metal lines and Kinman (1959)
reached a similar conclusion in a study
of integrated spectra of globular clusters. (Kinman's investigation was 
based on the spectral type 
obtained from a comparison of the G band with the H$\alpha$ line.)
As a result of
these findings, it is generally assumed that Oosterhoff type is 
associated with metal abundance.  
However, in their seminal paper on the Oosterhoff groups,
van Albada and Baker (1973) pointed out
that  the distribution of color on the horizontal branch was a more important
factor. They found that the fraction of HB stars on the blue side of the
RR Lyrae gap was greater in OoII clusters. Consequently,
they suggested that the RR Lyrae
variables in the OoII clusters must be evolving from blue to red through the
instability strip, while those
in the OoI clusters evolve from red to blue. This difference in direction
of evolution, combined with a hysteresis effect in the pulsations could
account for the difference in periods and
the different proportions of RRc stars between the
two groups. Later, Gratton, Tornamb\`e \& Ortolani (1986) and Lee, Demarque 
\& Zinn (1990, hereafter LDZ) 
pointed out that the RR Lyrae variables in the OoII clusters could be post ZAHB
stars whose ZAHB phase is on the blue side of the instability
strip. One would therefore expect the RR Lyrae variables in the OoII clusters
to evolve redward and to
have lower masses and higher luminosities than those in OoI clusters.

In Oosterhoff's original investigations,
$\omega$ Centauri was considered to belong to group II, the long period
group, but when 
Freeman \& Rodgers (1975) discovered that 
there was a diversity in composition among its RR Lyrae
variables, some investigators recognized that it had 
properties of both groups. 
Butler, Dickens \& Epps (1978) 
pointed out that the RR Lyrae variables with [Fe/H] greater than
$-1$ have the
characteristics of type I and that the more metal poor stars exhibit
type II characteristics. Taking a different approach,
Caputo \& Castellani (1975) demonstrated that the period-frequency plot
for the low luminosity RRab stars showed OoI characteristics and 
suggested that $\omega$ Cen provides a link between the two Oosterhoff
groups. 
The purpose of the present investigation is to  use pulsation theory to
derive relative masses 
for the $\omega$ Cen RR Lyrae variables 
to test the hypothesis that the OoII variables have lower masses and
higher luminosities than the OoI variables.
Our study is based on the observations made by the
OGLE team.

%

\section{THE OGLE DATA FOR RR LYRAE VARIABLES IN $\omega$ CENTAURI}

\subsection{Light curves, period-luminosity and period-amplitude relations}

In 1993, 1994 and 1995, CCD observations of $\omega$ Centauri
were made as a side-project of the OGLE (Optical Gravitational
Lensing Experiment) team. As a result, three papers on variable stars have
been published by Kaluzny and his collaborators
(Kaluzny \it et al. \rm 1996, 1997a, 1997b). The first two papers deal
with eclipsing binaries, SX PHe stars and spotted variables,
and the third, hereafter referred
to as K97b, presents periods and light curves for 140 RR Lyrae and population
II Cepheid variables. 

The present investigation is based on the $V$ magnitudes for
128 of the stars studied by K97b. 
Our sample consists of all of
the stars with mean V magnitude between
$14.0$ and $15.2$ and periods less than 0\day 9, with the  exception of 
OGLE\# 95, 96, 171 and 208 for which K97b
published periods close to one half day and 
amplitudes less than $0.2$ mag. 
The OGLE observations included seven different fields: 5139A, B, C, BC, D, 
E and F.  
Field 5139B, 
which covered the most central part of the cluster, 
contained more RR Lyrae variables than any of the other fields, but many of
these stars were included in other fields as well.
Since there are zero point shifts as large as $0.05$ mag from one field
to another, it was desirable to consider as many stars as possible in one
field only. Consequently, if a star was observed in field B, our analysis is
based only on the observations in field B.
Some stars were observed only in fields E and F, and in these
cases, we used the data from field E.  In general, we included
only magnitudes for which the listed
error was less than $0.02$ mag for our analysis. However, for
OGLE \#194, we included all magnitudes with errors less than $0.03$ in
order to obtain sufficient phase coverage..

To begin our analysis, we plotted  a light curve for each star using the period
listed in Table 1 of K97b. For some stars, we found small phase shifts in
the curves, 
and in these cases, we revised the periods. 
Our adopted periods are listed in Table 1 and light curves, arranged in 
order of increasing period, are shown in Figure 1. 
All but one of the periods
agree to within 0\day 0001 of  
those published by K97b. For star OGLE \#149, we found
a much shorter period, 0\day 378 compared with the K97b value 0\day 609. 
An examination of Fig.~1 shows how the general characteristics of the light
curves change with period. It was in a study of the RR Lyrae variables
in $\omega$ Cen that  Bailey (1902) set up his three subclasses
$a$, $b$ and $c$. 
He considered V8 (OGLE \# 199, P=0\day 5213) to be a 
typical star for subclass $a$, V3 (OGLE \#184, P=0\day 84126) 
for subclass $b$ and V35 (OGLE \#78, P=0\day 3868) for subclass $c$. 
The stars of subclass $c$ have the shortest periods and
type $b$ have the longest periods. 
In addition,
the light curves for the three subclasses differ in amplitude and 
in shape. 
It was subsequently 
demonstrated by Schwarzschild (1940), in a test of the period-density 
relation for RR Lyrae variables in M3, that the $c$-type variables
were most likely pulsating in the first overtone, while the $a$- and $b$-
types were pulsating in the fundamental mode. Since this mode difference
was recognized, the distinction between $a$ and $b$ has usually been 
disregarded and both groups have been referred to as RRab types.
However, because of the high quality of the light curves of Fig.~1, the
difference between types $a$ and $b$ is very clear. 
There is a sudden
change in slope about halfway  up the rising branch of the $b$ type
light curves. 
Other stars classified as $b$ by Bailey and subsequently by
Martin (1938) in his major study of $\omega$ Cen
are variables 15, 26, 34, 38, 54 and 85
(OGLE \# 124, 120, 90, 170, 74 and 176 respectively) and the light curves
for all of these stars show that same characteristic.

In Table 1, we list for each star the OGLE ID\#, the variable number from
Sawyer Hogg's (1973) catalogue,\footnote{ OGLE \# 191 and \#71 (V184 and V185)
were discovered by Butler \it et al. \rm (1978).}
the field, our adopted period, the $V$
amplitude, the mean $V$ magnitude and
the mode of pulsation according to the location of the star in the
period-amplitude diagram. The period-luminosity (P-L)
and period-amplitude (P-A) relations
are plotted in Fig.~2. The points in the P-A diagram seem
to fall into different regimes which we interprete to
be due to different modes of pulsation and have therefore
plotted a different symbol
for each mode.
Also, among the fundamental mode and second overtone pulsators,
there are two sequences in the P-A plot. A mean magnitude of $14.65$
roughly divides the stars into these two sequences. We  assume these occur
because stars of both Oosterhoff classes are present in $\omega$ Cen.
This will be discussed further in section 3.
In Table 1, the fundamental mode pulsators are denoted
as F and the overtone pulsators as 1st, 2nd or 3rd, depending on the mode of
pulsation. 
The amplitudes and mean magnitudes listed in Table 1 were obtained from fits of
the V magnitudes to a Fourier series of the form:
\begin{eqnarray}
mag = A_0+\sum _{j=1,n} A_j \cos (j\omega t + \phi _j)  
                     \hskip 2mm 
\end{eqnarray}
where $\omega$ is ($2\pi$/period).
For each star, the epoch was taken as HJD $2449000.000$ so that $t$ in the 
equation refers to (HJD-$2449000.000$) and HJD represents the heliocentric
Julian date of the observation.\footnote{OGLE \#71 and \#119 have periods
very close to 0\day 33 and as a result, the phase coverage was not complete.
However, for both of these stars, the maximum and minimum on the light
curve were well defined and so we estimated the amplitude from these.}
For the fundamental mode pulsators, the order of the fit
$n$ was 8 and for the overtone pulsators, it was 6. The adopted mean $V$
magnitude is 
$A_0$ from the fit of equation 1. 
In order to classify the light curves of the fundamental mode
pulsators as normal (n)  or peculiar (p), we used equations
derived by Kov\'acs \& Kanbur (1998) for testing the 
compatibilty condition of Jurcsik \& Kov\'acs  (1996, hereafter JK). 

The mean periods for the stars in our sample are as follows: 0\day 649 for the
59 fundamental mode, 0\day 383 for the 48 first overtone  and 0\day 304 for
the 21 second and third overtone pulsators. The  mean periods for the
fundamental and first overtone are very close
to those derived by van Agt and  Oosterhoff (1959) for the Oosterhoff type
II clusters. Thus, even though $\omega$ Cen seems to
contain RR Lyrae variables
of both Oosterhoff groups, most belong to group II.

\subsection{Evidence for Second Overtone Pulsation}

Although we have classified many of the RR Lyrae stars as
second overtone pulsators (RRe stars), the possibility of the existence of 
RRe stars
has been the  subject of some debate in the literature. 
A few years ago,
Stellingwerf, Gautschy \& Dickens (1987) calculated a model
for a second-overtone pulsator 
with an amplitude similar to first overtone pulsators. They predicted that
the light variation of an RRe star should have a sharper peak at maximum light.
However, they noted that they could not exclude the possibility that RRe stars 
may have lower amplitudes and  sinusoidal light variations.
The RRe candidates that we have identified in $\omega$ Cen have
sinusoidal light variations and low amplitudes compared to the
fundamental mode and first overtone pulsators.  This is illustrated
in Figure 3 where  we show the light curves for the three stars that Bailey
considered the prototypes for his subclasses $a$, $b$ and $c$ along with
the curve for OGLE \#191 (V184), one of the  stars we have classified as RRe.
There is a distinct 
progression in amplitude and degree of symmetry from subclass $a$ through
$c$ and \#191 forms a natural  extension to the sequence. 

Studies of P-A relations have led many other
investigators to recognize plausible
RRe candidates.
Evidence for an RRe star in the field (V2109 Cygni) has been presented 
by Kiss \it et al. \rm (1999). 
Credible RRe candidates have also
been identified in the globular clusters 
M68 (van Albada 
\& Baker 1973), NGC 4833 (Demers \& Wehlau 1977),
IC 4499 (Clement, Dickens \& Bingham 1979, Walker \&
Nemec 1996), M3 (Kaluzny \it et al. \rm 1998, hereafter K98)  and M5
(Kaluzny \it et al. \rm 2000, hereafter K2000). 
In addition, recent studies of M2 (Lee \& Carney 1999a) and NGC 5466
(Corwin, Carney \& Nifong 1999) indicate that 
there may be second overtone pulsators in these clusters as well.
However, Kov\'acs (1998a)
compared the RRe candidates in M68 and IC 4499 with the other RRc stars
and concluded that their relative
magnitudes and colors were incompatible with the assumption that they were
RRe stars. 
Furthermore, the models of Bono \it et al. \rm (1997, hereafter referred to
as B97) predict that  
a plot of amplitude versus period (or temperature) for first overtone
pulsators shows a characteristic `bell' shape with amplitudes decreasing
at shorter periods and higher temperatures. One would therefore expect
the first overtone variables with the shortest periods to have the
lowest amplitudes. This is exactly what Clement \& Shelton (1999a)
found in a recent study of the globular cluster M9; the P-A plot
for the overtone pulsators had the classic `bell'
shape. Thus, it is possible that some of
the short-period, low-amplitude variables in $\omega$ Cen are pulsating
in the first overtone mode.
Nevertheless, we believe that the shifts in the 
period-luminosity plot for the stars in the different period-amplitude regimes 
of Fig.~2 indicate 
the existence of more than two pulsation modes. 
Among the stars plotted in the 
upper panel of Fig.~2 with $<V>$ between $14.5$ and $14.65$, the 
RRe candidates (solid triangles) have $\log P$ in the range $-0.55$ to
$-0.45$ while the other RRc stars (open circles) have longer periods
($\log P$ between $-0.45$ and $-0.35$). The 
luminosities of both groups are comparable. 
RRe stars are expected to have shorter periods and lower amplitudes so
we might assume that the stars with shorter periods are RRe stars. On the other
hand, if they are all RRc stars, the stars with shorter periods must have
higher temperatures and/or higher masses. This is required by the 
period-density relation (see equation (7)). If the temperatures are the
same, the stars with shorter periods must have higher masses.
However, when we consider the amplitudes, it turns out that 
the short period stars can not have higher masses.
B97 made a number of plots to illustrate the dependence of
amplitude for first overtone pulsators on temperature, mass and luminosity. 
Their diagrams demonstrate
that, at constant temperature and luminosity, stars with lower amplitudes 
have lower not higher masses. This implies that if the solid triangles
represent first overtone pulsators, they must have higher temperatures
than the others. In fact, they must be hotter by $\Delta \log T_e\sim 0.029$
(almost 500K).
Another puzzling feature to note in the period-amplitude plot of Fig.~2 is the
general tendency for amplitude to decrease with increasing period for the
solid triangles brighter than $<V>=14.65$. This is not the characteristic
`bell' shape predicted by B97 for period amplitude plots of first
overtone variables.
We therefore conclude that 
these stars are  probably not first overtone pulsators; rather they  
are pulsating in the second overtone mode. 
There may even be third 
overtone pulsators among the RR Lyrae population in $\omega$ Cen!

Most of the second and third overtone pulsators listed in Table 1 have no
Sawyer Hogg numbers.  
This is because their amplitudes are generally lower than those of the RRc 
stars and as a result, they were not identified in the
earlier photographic studies of $\omega$ Cen
by Bailey (1902) and Martin (1938). 
It is also possible that the four stars that we exluded from our investigation,
OGLE \#95, 96, 171 and 208 are second overtone pulsators.

An independent argument for the existence of RRe stars has been proposed by
the MACHO consortium.
In a study of RR Lyrae variables in the LMC, Alcock \it et al. \rm (1996)  
showed that the period frequency distribution had three peaks, at 
periods of 0\day 58, 0\day 34, and 0\day 28 respectively. They
interpreted this to mean there were three pulsation modes. 
In Fig.~4, we plot a period-frequency distribution for the 
OGLE sample of $\omega$ Cen RR Lyrae variables. It also has
three peaks, but the periods associated with the peaks are longer than those
in the LMC. They occur at periods of 0\day 63,
0\day 40 and 0\day 32. 
We assume that the periods for the
peaks differ because most of the RR Lyrae variables in $\omega$ Cen have
OoII characteristics, while the LMC has mainly an OoI population. 
An examination of Figure 2 shows that, although there
are some RRc stars with periods of approximately 0\day 32 ($\log P=-0.49$),
the peak would
not occur if there were no RRe stars. 
We therefore concur with Alcock \it et al. \rm that 
the reason for the extra peak at short periods in the period-frequency 
distribution of both
systems is the presence of RRe stars.
There is another important difference between the 
period-frequency distributions of $\omega$ Cen and the LMC -- the size
of the peaks. For $\omega$ Cen, they are all at approximately the same
height, but for the LMC, the height of the peak increases with period,
thus
implying there are more RRab stars than RRc stars and more RRc stars than RRe
stars. 
It is well known that OoI systems have more RRab stars than RRc stars and
van Albada and Baker (1973) attributed this to different predominant
directions of 
evolution through the instability strip for the two groups, accompanied by
a hysteresis effect in the pulsation.
This  could account for the fact that the LMC has significantly
fewer RRe stars than RRc stars. However, it may also be a detection problem
because the RRe stars have lower amplitudes.

\subsection{Evidence for Non-Radial Pulsations}

With accurate CCD photometry, it is possible to identify low level changes in
the amplitudes of RR Lyrae variables and a number of such
changes have been noted. 
Walker (1994) found short term
variations in the light curve of the short period RRc (RRe candidate)
star V5 in M68.  More recently,
Olech and his collaborators 
found changes on time scales of a few days in the
light curves of three RRc stars in M55 (Olech \it et al. \rm 1999a) 
and one in M5 (Olech \it et al. \rm 1999b). They also
found that these stars all had multiple periods. For example, in
their analysis of V104 in M5,  they identified 
two clear peaks in the periodogram at periods of
0\day 332 and 0\day 311 and they attributed these to the presence of
non-radial oscillations, similar to those observed in many $\delta$ Scuti
stars.  Theoretical calculations made by Van Hoolst \it et al. \rm (1998) have
demonstrated that non-radial modes can be excited in RR Lyrae variables as
well. Non-radial oscillations have also been detected in some
RR Lyrae stars in the Galactic Bulge
by Moskalik (2000) and in the LMC by Kov\'acs \it et al. \rm (2000).

The OGLE dataset for $\omega$ Cen includes a number of RR Lyrae stars that
change on a time scale of days. In particular, K97b commented on
the instability of the light curve of \#162, but an examination of
the light curves of Fig.~1 indicates that there are
others as well.  In Fig.~5, we show a light curve for \#186, based on 10 
consecutive nights in 1995. The observations for each night are
plotted with a different symbol
and it is very clear that there are changes on the scale of a few days.
In Fig.~6, we present $\Theta$-transforms for \#186; the real data are plotted
in the upper panel and the prewhitened data in the lower panel. The diagram 
indicates that
there are two well defined periods, just as Olech \it et al. \rm found for
the stars that they studied. Thus we conclude 
that \#186 is an RRe star that exhibits non-radial pulsations.

\subsection{Physical Properties derived from Fourier parameters}

\subsubsection {RRc stars}

In the last few years, two methods have been devised for deriving
luminosities of RR Lyrae variables 
from Fourier parameters. 
Since there is a range of more than $0.5$ in the apparent $V$ magnitude
among the RR Lyrae variables in $\omega$ Cen, it provides a good sample
for testing these methods. 

In an investigation of hydrodynamic 
pulsation models for RRc stars, Simon \& Clement (1993a,b) 
analysed the light curves and
derived the following relationships:
\begin{eqnarray}
\log L/\lsun=1.04\log P -0.058\phi_{31} +2.41  
\end{eqnarray}
\begin{eqnarray}
\log M/\msun = 0.52 \log P -0.11\phi_{31}+0.39  
\end{eqnarray}
\noindent where
$L/\lsun$ is the luminosity in terms of the Sun's luminosity,
$M/\msun$ is the mass in solar units,  
$P$ is the pulsation period in days,
$\phi_{31}$ is the Fourier phase difference 
($\phi_3-3\phi_1$) computed from the fit of equation (1) to the points.
(The standard deviations of the fits for $\log L/\lsun$ and $\log M/\msun$ to
equations (2) and (3)
were $\sigma=0.025$ and $0.03$ respectively.)
Thus the analysis indicated that $\phi_{31}$ depends essentially
on mass and luminosity and that it is insensitive to metallicity. This
means that equations (2) and (3) can be applied to stars with different
metal abundance.
Once the mass and luminosity have been evaluated, the effective
temperature can also be calculated from the period-density relation
which can be expressed in the following form:
\begin{eqnarray}
\log T_e= 3.265 -0.3026 \log P - 0.1777\log M/\msun + 0.2402 \log L/\lsun
\end{eqnarray}
Kov\'acs and his collaborators took a different approach. They derived 
empirical 
formulae for calculating $M_V$ by studying systems in which there are large 
numbers of
RR Lyrae variables, e.~g. the Sculptor dwarf galaxy. 
For the RRc stars, Kov\'acs (1998b) derived the following formula for
calculating the absolute magnitude:
\begin{eqnarray}
M_V=1.261-0.961 P -0.044\phi_{21} -4.447A_4
\end{eqnarray}
where $P$ is the period, $\phi_{21}$ is the phase difference
$(\phi_2-2\phi_1)$ based on a sine series fit,
$A_4$ is the fourth order amplitude and the standard deviation of the fit
was $0.042$.
The zero point was based on the 
Baade-Wesselink luminosity scale of Clementini \it et al. \rm (1995). 
It should be noted that the luminosity scale for RR Lyrae variables
is still the subject of some controversy, but even if there is an 
error in the zero point for 
equation (5), the relative ranking of the magnitudes will not
be affected.

In Table 2, we list 
the Fourier phase differences $\phi_{21}$, $\phi_{31}$, (along with
their errors)  and the amplitude $A_4$ for all of the overtone pulsators 
derived from equation (1). 
Also included are $M_V$, $\log L/\lsun$ and $M/\msun$ 
calculated from equations (2), (3) and  (5) and [Fe/H]$_{hk}$ values recently
determined by Rey \it et al. \rm (2000, hereafter R2000) from C$aby$ 
photometry.  The stars are arranged in
order of increasing period. The asterisks beside some of the star numbers
refer to the Oosterhoff type and will be discussed further in section $3.1$.
In our calculations of
$M_V$, $\log L/\lsun$ and $M/\msun$, 
we included only the stars
for which the error in the phase difference is less than $0.2$.
For most of the short-period stars, 
the errors are large, particularly those for $\phi_{31}$. This is
because of their low amplitudes and sinusoidal light curves. 
Kov\'acs'
Fourier fit was based on a sine series, but the parameters that
we list in Table 2 are based on a cosine series.  Therefore,
to put our values for the phase differences on the same 
system as his, we subtracted $1.57$ from $\phi_{21}$ before we computed $M_V$. 
In Fig.~7, we plot the computed $\log L$ and $M_V$ against the mean $V$ 
magnitude. The first overtone pulsators are plotted as solid circles and
the second overtone pulsators as open triangles. The crosses represent
stars not considered to be cluster members. The open circles enclosed in boxes
denote four anomalous variables \#132, \#169 (V123), \#185 (V47) and \#73 
(V68)  which all have unusually high values of
$\phi_{21}$ according to Table 2. 
We have classified them as RRc stars even though they all 
have periods greater than 0\day 44.  
The envelope lines in the upper panel
are separated by $\Delta V=0.12$, the uncertainty in the calibration
of equation (2) and most of the solid circles lie either on these
lines or between them. 
It appears that equation (2) is effective for predicting
the relative luminosities of the RRc stars.
The envelope lines in the lower panel are separated by $0.085$, the uncertainty
in Kov\'acs' (1998b) calibration and here the fit is not quite so good. 
For the 
faintest first overtone pulsators, the predicted magnitudes are too bright.
On the other hand, equation (5) predicts fainter
values of $M_V$ for the second overtone
pulsators due to their shorter periods and lower $A_4$. 
No values of $\log L$ were
computed for second overtone pulsators because the errors in their $\phi_{31}$
values
were all greater than our threshold $0.2$ because of their low amplitudes. 

In Fig.~8, we show the relationship between $\log L$ and
$M_V$ calculated from the Fourier parameters.  They correlate well and this
leads us to conclude that the two outliers plotted as crosses
in Fig.~7 at $M_V\sim 0.8$ (\#97 and \#192) are
not cluster members. They just happen to be in the field of view. Another
star that is probably not a member is \#181 which, with $<V>=15.181$, is off
the scale of Fig.~7. However, it is plotted in Fig.~8 and lies between the 
envelope lines.  The outlier at $M_V=0.477, \log  L=1.77$ in Fig.~8 is 
star \# 185
(P=0\day 485), one of the anomalous stars with large $\phi_{21}$. The 
light curve for this star has an inflection on
the rising branch that is different from most of the other RRc stars.
The light curve for \#169 is similar, but it
is not plotted in Fig.~8 because its $\phi_{31}$ error is greater than $0.2$.
If it were, the
point would also lie  to the right of the envelope lines.

\subsubsection {RRab stars}

An empirical formula for relating the absolute 
magnitude to period and Fourier coefficients for RRab stars was derived by
Kov\'acs \& Jurcsik (1996): 
\begin{eqnarray}
M_V=1.221-1.396 P-0.477A_1 + 0.103\phi_{31}
\end{eqnarray}
In Table 3, we list the Fourier amplitude $A_1$, phase difference
$\phi_{31}$, $M_V$ calculated from equation  (6) for the RRab 
variables with `normal' light curves and the [Fe/H]$_{hk}$
values recently determined by R2000. The stars are listed in order of
increasing period.
The Fourier coefficients in equation (6) are based on a sine series.
Therefore, to
put our coefficients from Table 3 on the appropriate system, we added
$3.14$ to $\phi_{31}$ before calculating $M_V$.  In Fig.~9, we plot $M_V$
against $<V>$ and the two quantities correlate reasonably well. However, the
fit for the $\omega$ Cen RRab stars is not as good  as the fits for
NGC 6171 and M3 based on the observations of
Clement \& Shelton (1997) and Kaluzny \it et al \rm (1998). 
The envelope 
lines have a slope of unity and their separation $\Delta V=0.1$ represents 
the uncertainty in the values of $M_V$ derived from equation (6). Most
of the points fall between the two lines, but \#174 (V84) with $<V>=14.291$
is displaced by more than $0.3$ mag. Bailey did not determine a period for
this star, but in Fig.~1, it is clear that its light curve has the classic 
`$b$' characteristics
in spite of its relatively short period (0\day 5799). In
their paper on the chemical inhomogeneity of $\omega$ Centauri,
Freeman
\& Rodgers (1975) found that the strength of the Ca II K-line for this star was 
greater than that of any of the other 27 RR Lyrae variables in their sample. 
We therefore conclude that \#174 (V84) is a foreground
star that belongs to a different population.
%
%

\section{THE OOSTERHOFF DICHOTOMY IN $\omega$ CENTAURI}
\subsection{The Period-Amplitude Relation}

In a study of the period-amplitude relation of RRab stars in the OoI
globular clusters M107, M4, M5, M3, and the OoII clusters M9, M68 and M92,
Clement \& Shelton (1999b) found that the period-amplitude relation for RRab 
stars appears to be  a function of 
Oosterhoff type. They therefore concluded that evolutionary effects were more
important than metal abundance for determining where a star lies in the
period-amplitude plane provided the star had a `normal' light curve.
In Fig.~10, 
we plot $A_V$ versus $\log P$ for the $\omega$ Cen
fundamental mode pulsators with `normal' light curves
and for first overtone stars with
errors less than $0.2$  in $\phi_{31}$. 
Only stars 
that are considered to be members of the $\omega$ Cen population 
according to the discussion in sections $2.4.1$ and $2.4.2$
are included in the plot.
Stars with $<V> \le 14.65$
are plotted as solid triangles or solid circles and the fainter
stars are plotted as open triangles or open circles. The straight line
between $\log P=-0.24$ and $-0.05$ is a least squares fit to the RRab stars
with $<V> \le 14.65$ and its location in the P-A plane is close to the 
line that Clement \& Shelton (1999b) derived for OoII RRab stars.
Most of the (solid) points lie very close to this 
line; their mean deviation $<\Delta \log P>$ is $0.014$. 
Their mean period is 0\day 673, a value appropriate for an OoII 
classification.
Jurcsik (1998a) has previously pointed out that the bulk of
the RRab stars in $\omega$ Cen comprise a very homogeneous group
and one can see this in Fig.~10.
The straight line between $\log P=-0.32$ and $-0.14$ is the OoI P-A
relation that Clement \& Shelton (1999b) derived 
from the M3 data of K98.
Several of the fainter RRab stars (the open circles and the open
triangles) lie close to this OoI relation. 
The mean period for the fainter stars is 0\day 542, a typical OoI value.
Thus the  P-A relation for the RRab seems to be effective for distinguishing
between Oosterhoff types.  
The fainter RRab stars are more metal
rich on the average; their mean [Fe/H] is $-1.53$ compared with $-1.74$
for the stars with $<V>\le 14.65$. However, there is not a strong
correlation between $<V>$ and [Fe/H]. This point has already been made
by R2000 and by Demarque \it et al. \rm (2000)
who note that the luminosity also depends on evolutionary status or HB
morphology. 

The other lines in Fig.~10 represent
least squares fits to the P-A relations for the RRc stars in M3 (OoI)
and M68 (OoII)
based on the data of K98 and Walker (1994).
In this case, the distinction  between OoI and II is not so marked for
stars brighter or fainter than $14.65$. It seems that, for the RRc stars,
the $<V>=14.65$ cutoff 
distinguishing between OoI and OoII types is too faint.
Therefore, in our discussion of RRc stars in the next section, we 
decide Oosterhoff type according to position in the P-A diagram. RRc stars
with periods less than 0\day 35 lie closer to the M3 line and so we consider
them to be OoI variables.
Stars with periods in the range 0\day 35 to 0\day 43 are considered to be
OoII variables. The OoI stars 
are indicated by single asterisks and the OoII stars by double asterisks in
Table 2.

A curious feature of Fig.~10 is the existence of first overtone
pulsators with periods in the range of 0\day 44 to 0\day 54. 
M68 does not have any overtone pulsators with
such long periods, even among the double-mode pulsators whose first overtone
periods range from 0\day 39 to 0\day 41.
These long period stars in $\omega$ Cen
are brighter than the others and as we noted in section $2.4$, they
have unusually
large values for $\phi_{21}$. The stars with the inflection on the
rising branch of the light curves (\#169 and \#185) are among them. 
The recent study of M5 by K2000 shows that M5-V76 has similar properties.

\subsection{The Masses and Luminosities of the RRc stars}

In section $2.4.1$, we showed
that equation (2), which was derived from hydrodynamic pulsation models by
Simon \& Clement (1993a),
successfully ranks the relative
luminosities of the $\omega$ Cen RRc stars.  
K2000 reached the same conclusion  in their recent study of M5.
We now use the masses computed from equation (3), which was also derived
from the models, to compare the masses and luminosites of RRc stars
in clusters belonging to both Oosterhoff groups.
Our results are summarized in 
Table 4 where we compare the data for $\omega$ Cen with six other well studied
globular clusters, three from each Oosterhoff group. 
These are M107 (Clement \& Shelton 1997), M5 (K2000), M3 (K98), 
M55 (Olech \it et al. \rm 1999a), M68 (Walker 1994) and M15 (Clement \& Shelton
1996, based on the data of Silbermann \& Smith 1995). 
For each cluster, we list [Fe/H], the number of stars analysed,
their mean period, mean mass,  mean luminosity and mean temperature.
For these last three quantities, we also include the standard error of the 
mean.
Most of the [Fe/H] values are taken from the compilation of Zinn (1985), but
for $\omega$ Cen, they are means of the [Fe/H]$_{hk}$ values derived for the
individual stars by R2000.
The $\omega$ Cen variables are segregated according to Oosterhoff type. 
The four bright stars with periods greater 
than 0\day 44 are not included. Their derived masses  are
uncertain because they are well below the range of the models
which were computed for masses ranging from $0.55$ to $0.85\msun$.
The clusters are arranged in order of increasing $\log L/\lsun$ which is,
in general, the order of decreasing metal abundance and decreasing temperature. 
The mean period increases through the sequence until M68 and M15, the 
two OoII clusters that have the  highest frequency of double-mode RR Lyrae
variables (RRd stars).\footnote{Recent CCD 
observations of the RRd stars in M68 and M15 
have been published by Walker (1994) and Purdue \it et al. \rm (1995)
respectively.}
In these two clusters, the overtone pulsators with periods longer than 0\day 39
are RRd stars and as a result,
the mean periods for their RRc stars are shorter than those for M55.
The masses calculated from equation (3) are low compared with those derived
from evolutionary models (Dorman 1992) and other investigations of pulsation.
However, the relative ranking is in agreement with masses derived
from the analysis of RRd stars. Studies of pulsation models
by Kov\'acs \it et al. \rm (1992), Cox (1995) and Bono \it et al. \rm (1996)
all indicate that the RRd variables in the OoII clusters M15 and M68
are more massive than those in the OoI clusters M3 and IC 4499. Most of
their models indicate that the difference
between the two is approximately $0.1\msun$. Three of these clusters are
included in Table 4 which indicates that the RRc stars in M15 and M68 are
$0.13\msun$ more massive than the ones in M3.
Another point to note about M68 and M15 is that the masses 
derived for their RRc stars from $\phi_{31}$ are considerably higher than
those for $\omega$ Cen and M55. 
If this is correct, then the data 
of Table 4 suggest that the double-mode phenomenon may occur only in
higher mass stars.  Recent studies of M55 and $\omega$ Cen have not
revealed any RRd stars.  Furthermore,
Nemec, Nemec \& Norris (1986) analysed Martin's (1938) published 
observations of the $\omega$ Cen RR Lyrae variables and found no RRd stars. 
Freyhammer, Andersen \& Petersen (1998, 2000) have detected
multi-mode pulsation in a few variables in $\omega$ Cen, but they are SX Phe, 
not RR Lyrae variables.

Another feature of the sequence of masses listed in
Table 4 is the discontinuity 
at the interface between the two Oosterhoff groups.
Although there is an increase of mass with luminosity among the clusters
in each group, there is a sudden drop in mass at the transition. 
The OoII RRc stars in $\omega$ Cen and M55 have lower  masses and higher
luminosities than the OoI RRc stars in $\omega$ Cen and M3.
This is exactly what is predicted by LDZ. The
OoII RR Lyrae variables are evolved stars 
from the blue horizontal branch so that
when they cross the instability strip, they have lower masses and higher 
luminosities than the OoI variables.

\subsection{The Masses of the RRab stars}

We now use the period-density relation to compare the 
masses of the RRab stars of the two
Oosterhoff groups in $\omega$ Cen.
The period-density relation can be expressed as a relationship among pulsation 
period, luminosity, mass and surface effective temperature. 
From models based on OPAL opacities, Cox (1995) 
derived a relation for fundamental mode RR Lyrae
pulsators with composition comparable to the $\omega$ Cen 
variables (Z=$0.0003$):
\begin{eqnarray}
\log P=11.519 + 0.829 \log L/\lsun -0.647 \log M/\msun -3.479 \log T_e
\end{eqnarray}
Our procedure is to select pairs of stars for which the amplitude is the
same (within $0.02$ mag), one from each Oosterhoff group
according to the P-A diagram of Fig.~10,
and then use the period-density relation to calculate the difference in mass.
From equation (7), we derive:
\begin{eqnarray}
0.647 \Delta \log M/\msun =
-0.332 \Delta m_{bol} -  \Delta \log P  - 3.479 \Delta \log T_e
\end{eqnarray}
where $m_{bol}$ refers to the apparent bolometric magnitude. 
We limit our selections to pairs of stars in the same
field so that the luminosity  difference obtained from the observations
is reliable. Otherwise the shifts in zero point between the different 
fields would introduce large uncertainties into the results. 

In order to apply equation (8), we need to know the temperature differences,
but the temperatures of the stars are not known. 
Our approach therefore is to evaluate 
$\Delta \log M/\msun$ for three different assumed values of $\Delta \log T_e$,
selected in the following manner.
We choose our first value $\Delta \log T_e=0.000$ because
B97 showed that the correlation between amplitude
and effective temperature of fundamental mode pulsators does not have
a strong dependence on mass or luminosity. Their study was based on
nonlinear pulsation models with helium content $Y=0.24$. On this basis,
one would expect a one-to-one correspondence between amplitude and
temperature. Our second value for $\Delta \log T_e$ is $-0.004$. We choose
this value because
Sandage (1993) presented evidence that OoII RRab stars are cooler than
those of OoI. For RRab stars at the blue fundamental edge of the instability
strip, he derived the relationship: $\Delta \log T_e=0.012 \Delta[Fe/H]$.
Among the star pairs we have selected, the OoII RRab stars are more metal
poor by an average amount $-0.34$. Therefore, based on Sandage's relationship,
we derive a mean $\Delta \log T_e=-0.004$.
We choose our third value $-0.005$ because our
analysis of the RRc stars in $\omega$ Cen also indicates that the
OoII RR Lyrae variables are cooler.
According to the data of Table 4, the mean temperature of the OoII RRc 
stars in $\omega$ Cen is lower by $\Delta \log T_e=0.005$ than that of the 
OoI stars. 
For equation (8), we also need to know the bolometric corrections
(BC) in order to compute $m_{bol}$. 
Jurcsik (1998b) derived equations for relating BC
to [Fe/H] and temperature and  from her equations,
we derive the following relationship:
\begin{eqnarray}
\Delta BC = 1.56115 \Delta \log T_e + 0.0445 \Delta [Fe/H].
\end{eqnarray}
We use equation (9) to calculate $\Delta BC$ for each assumed
value of $\Delta \log T_e$.

Our selected pairs of stars (seven in all)
are listed in Table 5. Also listed are  the observed
differences $\Delta \log P$, $\Delta V$, $\Delta A_V$ and $\Delta [Fe/H]$
and the $\Delta \log M$ values calculated from equation (8) according to
the different assumptions for $\Delta \log T_e$. On the last line, the
mean value for each quantity is listed. For each assumed temperature
difference, the mean mass of the OoII stars is lower 
than the mean for the OoI stars, just as we
found for the RRc stars. This is further support for the LDZ hypothesis.
If the LDZ hypothesis
is correct, one would also expect to find that the periods
of OoII RR Lyrae stars increase. This is exactly what has been found
in a recent study of the period changes of RRab stars in $\omega$ Cen
by Jurcsik (2000) and Jurcsik \it et al. \rm (2000).

Although our analysis of the RRab star pairs indicates that, in general
masses are lower and luminosities are higher for the OoII stars, there are
individual cases that are exceptions. For example, the OoII star \#147
has a higher mass than the OoI star \#154 for all of our assumed values of
$\Delta \log Te$. This is because \#147 is so much brighter than
\#154 that its mass must also be larger to account for its observed period.
Another anomalous star is \#113. For six of the seven pairs listed in Table 5,
the OoII star is brighter, but the OoI star \#113 is brighter than its OoII
counterpart. In fact, it is brighter than any of the other stars that lie near
our adopted OoI P-A relation. An examination of Fig.~1 shows that its light 
curve does not have the characteristic bump and dip preceding the rise to
maximum light. Theoretical light curves published by B97
illustrate that, at constant mass and temperature, models with higher
luminosity have a less pronounced bump. This is exactly what we observe in
the light curve of \#113. It may have a mass and temperature similar to the 
other OoI stars, but it is more luminous. With high quality light curves
based on CCD observations, it is becoming possible to distinguish between
different models.

\section{The Origin of the Dichotomy}

Our analysis of the RR Lyrae stars
in $\omega$ Cen has confirmed that the cluster contains variables with the
properties of both Oosterhoff groups. 
We have also presented evidence to illustrate that both the RRc and RRab
OoII variables belong to a more evolved population than the OoI
variables. On the average, the OoII variables are less massive, more luminous
stars that have spent their ZAHB phase on the blue side of the instability
strip. 
The same conclusion can be reached from a comparison of the first
overtone pulsators in the OoI cluster M3 and the OoII cluster M55. 
One explanation for this advanced evolutionary state is that their ZAHB
masses are low because of 
a high mass-loss rate at the tip of the red giant branch.
Alternatively, it could be that the masses of their
main sequence progenitors are lower and consequently,
the OoII variables are older.
This is the conclusion that Lee \& Carney (1999b) reached 
in their investigation of M2 and M3,
two clusters that have similar metallicities, but different Oosterhoff types. 
By comparing the difference in color between the
base of the giant branch and the main sequence turnoff, they estimated that
the OoII cluster M2 is about 2 Gyr older than the OoI cluster M3. 
It is possible that the Oosterhoff dichotomy
in $\omega$ Cen is also due to an age difference. Recent studies of
$\omega$ Cen by Hilker \& Richtler (1999) and 
Hughes \& Wallerstein (2000) present 
evidence for a range in age of at least 3 Gyr. Both groups performed
Str\"omgren photometry 
(Hilker \& Richtler 
 observed red giants and 
Hughes \& Wallerstein  observed stars
near the main sequence turnoff) and both found that the more metal rich stars
are at least 3 Gyr younger. 
The data listed in Tables 4 and  5 of the present 
paper indicate that the OoI RR Lyrae variables in $\omega$ Cen are more
metal rich than the OoII variables. Thus if the age-metallicity correlation
can be extended to HB stars, a difference in age may be the
cause of the Oosterhoff dichotomy in $\omega$ Cen.
An examination of Dorman's (1992) oxygen-enhanced 
models for HB stars
with [Fe/H]=$-1.48$ and $-1.66$ (metallicities comparable to those
observed in the $\omega$ Cen RR Lyraes) shows that at $\log T_e= 3.86$ 
(a temperature appropriate for the instability strip), there are models with
[Fe/H] = $-1.66$ and $\log L= 1.74$ that are less massive than models with
[Fe/H]=-1.48. However, the mass difference between the two is less than
$0.01\msun$. The mass differences listed in Tables 4 and 5 are greater. 
Clearly, there is still more work to be done before we can fully understand
the evolution of RR Lyrae variables and the Oosterhoff dichotomy.

%
%
\acknowledgements

We would like to thank Giuseppe Bono, Bruce Carney,
Marcio Catelan, Johanna Jurcsik, Andy Layden, Geza Kov\'acs, Joergen Petersen,
Robert Rood, Horace Smith and Allen
Sweigert 
for discussing this work during the course of the investigation. 
Support for this work from
the Natural Sciences and Engineering Research Council of Canada is
gratefully acknowledged.

\newpage

\begin{figure}
\plottwo{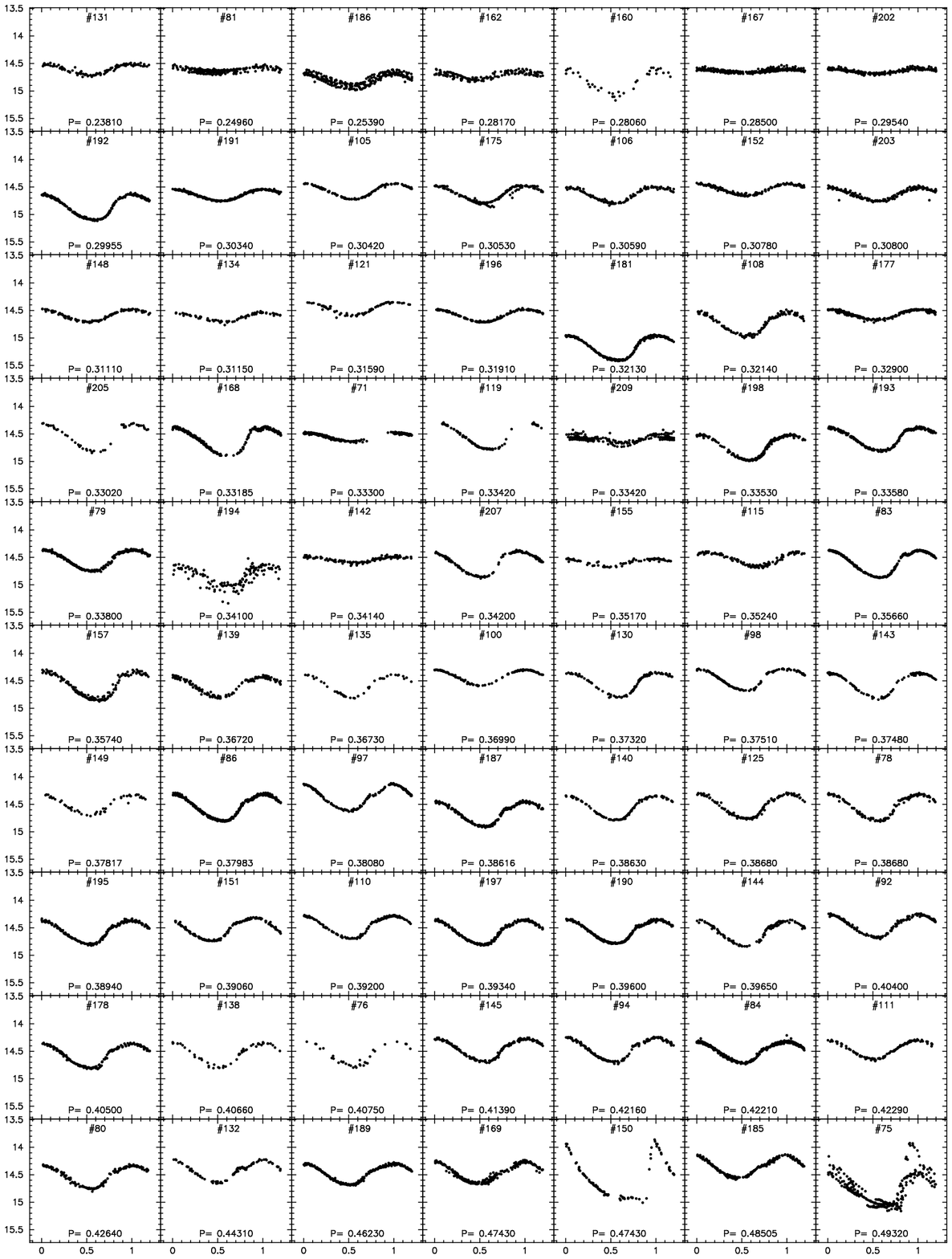}{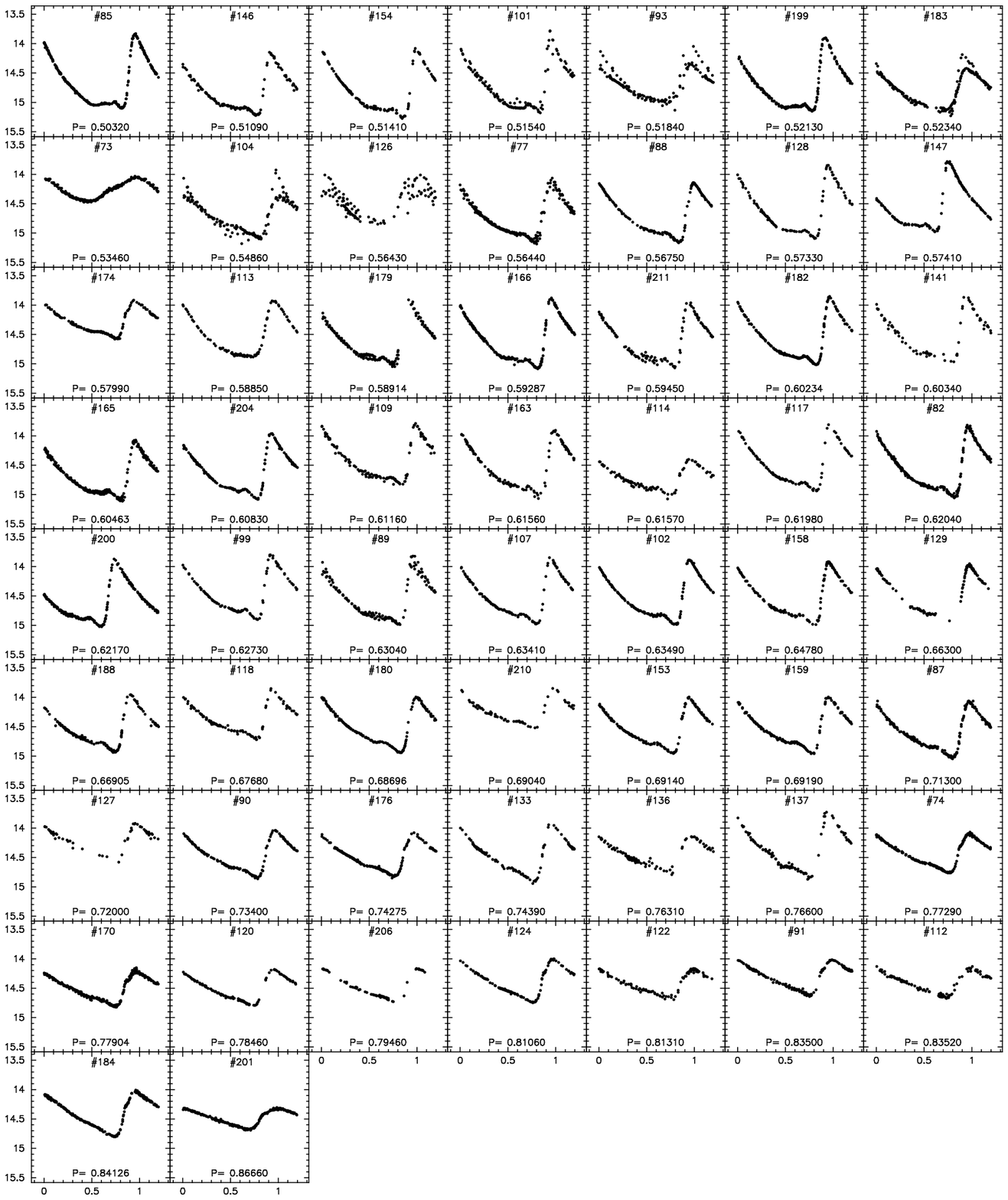}
\caption
{$V$ light curves for the 128 stars in our sample. The curves are arranged
in order of increasing period.}
\end{figure}

\begin{figure}
\plotfiddle{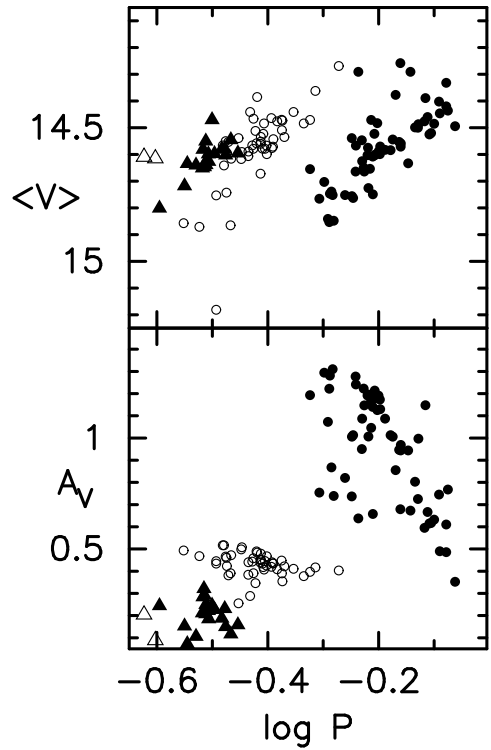}{6.0 truein}{0}{80}{80}{-200}{-10}
\caption
{The period-luminosity and  period-amplitude 
relation ($<V>$ and $A_V$ versus $\log P$) for the RR Lyrae variables in 
$\omega$ Cen. The points in the diagram fall into four different regimes
which appear to correspond to different modes of pulsation and so we have
plotted them with different symbols. The solid circles
represent the fundamental mode, the open circles denote the first overtone, 
the solid triangles are  the second overtone and the open triangles 
are the third overtone.
Among the fundamental mode and second overtone pulsators, the P-A relation
appears to be segregated into two groups according to luminosity. This occurs
because stars of both Oosterhoff types are present in $\omega$ Cen.}
\end {figure}

\begin{figure}
\plotfiddle{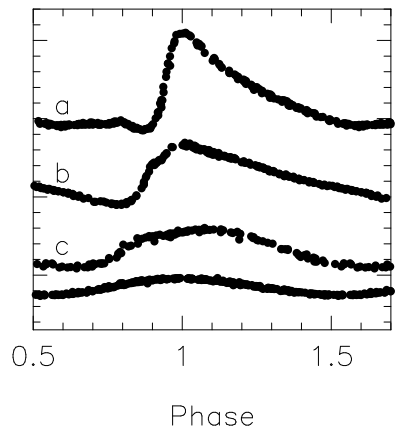}{6.0 truein}{0}{80}{80}{-200}{-10}
\caption
{OGLE $V$ light curves for the RR Lyrae variables Bailey (1902) considered to 
be the prototypes for his three subclasses. These are
\#199, (type $a$, P=0\day 5213), \#184 (type $b$,
P=0\day 84126) and \#78 (type $c$, P=0.3868). Bailey type $a$ and $b$ variables
pulsate in the fundamental mode, while type $c$ variables pulsate in the
first overtone.
The bottom curve in the
diagram is for \#191 (P=0\day 3034). 
There is a marked decrease in amplitude and increase in symmetry of the curves
as one progresses through the Bailey subclasses and \#191 is an extension of 
the sequence.
We consider it to be a second overtone pulsator.}
\end {figure}

\begin{figure}
\plotfiddle{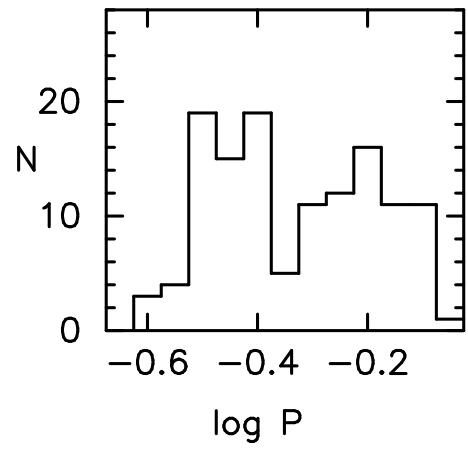}{6.0 truein}{0}{80}{80}{-200}{-10}
\caption
{Period-frequency distribution for the RR Lyrae variables in the OGLE sample
for $\omega$ Cen. The peak at $\log P=-0.5$ is due in part to the
presence of second overtone pulsators (RRe stars).}
\end {figure}

\begin{figure}
\plotfiddle{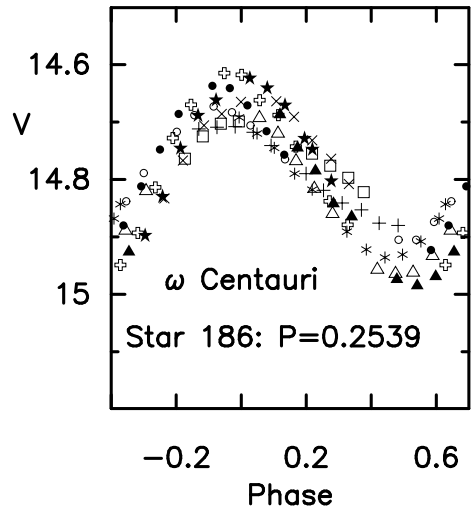}{6.0 truein}{0}{80}{80}{-200}{-10}
\caption
{$V$ light curve for OGLE \#186 on 10 consecutive nights in 1995 (JD 2449821
to 2449830). Each night is plotted as a different symbol illustrating
that the amplitude changes from cycle to cycle.}
\end {figure}

\begin{figure}
\plotfiddle{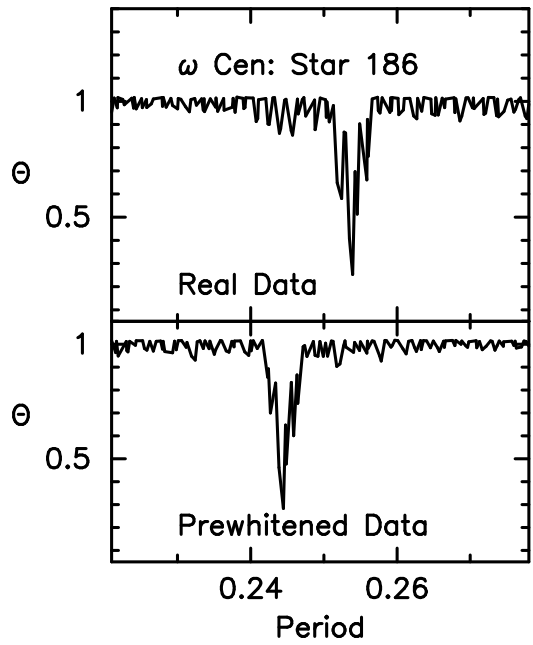}{6.0 truein}{0}{80}{80}{-200}{-10}
\caption
{$\Theta$-transforms for the real and prewhitened data for
OGLE \#186. In the upper panel, the minimum occurs at 
0\day 2539, the adopted period. After prewhitening with this period,
the  minimum $\Theta$ value 
is 0\day 2445. This multiple periodicity is probably due to non-radial 
oscillations.}
\end {figure}

\begin{figure}
\plotfiddle{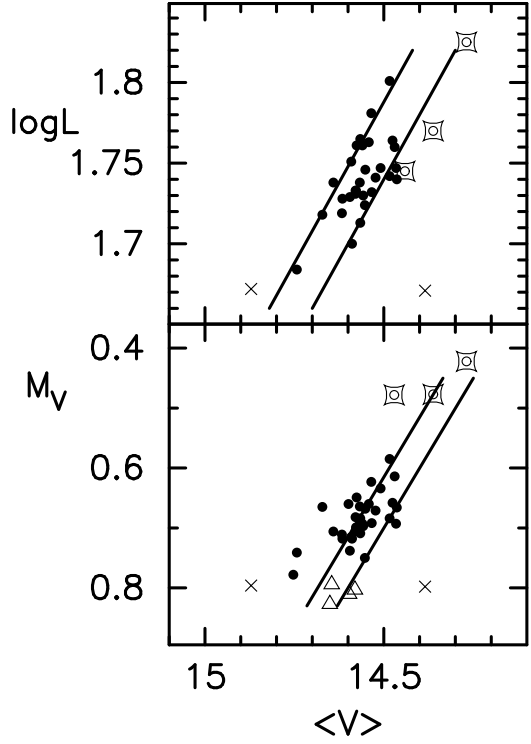}{6.0 truein}{0}{80}{80}{-200}{-10}
\caption
{$\log L$ derived from equation (2) and $M_V$ derived from equation (5) plotted
against $<V>$ for RRc stars (solid circles) and RRe stars
(open triangles) in $\omega$ Cen. The open circles enclosed in boxes
represent 4 stars that have anomalously high values of $\phi_{21}$. Stars
considered to be non-members are plotted as crosses.
The envelope lines in the upper panel have a slope of $0.4$ and are separated
by $\Delta V=0.12$ which represents the uncertainty of the fit of equation
(2) to the models. The lines in the lower panel
have a slope of unity and
are separated by $\Delta V=0.085$ the uncertainty in the calibration of 
equation (5).}
\end {figure}

\begin{figure}
\plotfiddle{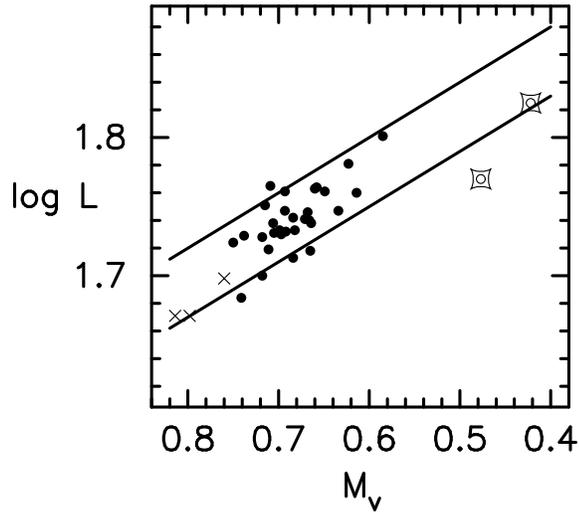}{6.0 truein}{0}{80}{80}{-200}{-10}
\caption
{$\log L/\lsun$ derived from equation (2) for the RRc stars in $\omega$ Cen.
plotted against 
$M_V$ derived from equation (5). 
The symbols are the same as in Fig.~7.
The envelope lines have a slope of $0.4$ and are separated by
$\log L =0.05$ which represents the uncertainty in the fit of the models
to equation (2).}
\end {figure}

\begin{figure}
\plotfiddle{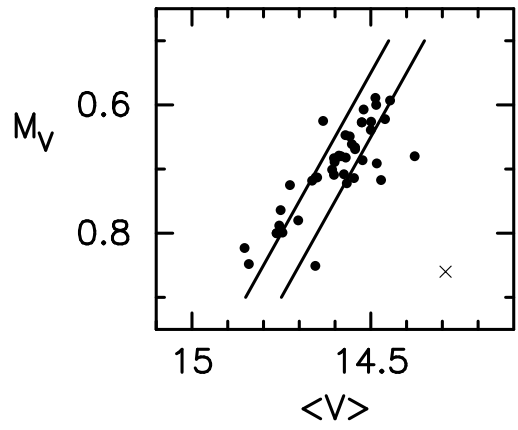}{6.0 truein}{0}{80}{80}{-200}{-10}
\caption
{$M_V$ calculated from equation (6) plotted against $<V>$ for the RRab stars
with `normal' light curves. 
The cross represents \#174  (V84) which is considered to be a non-member.}
\end {figure}

\newpage

\begin{figure}
\plotfiddle{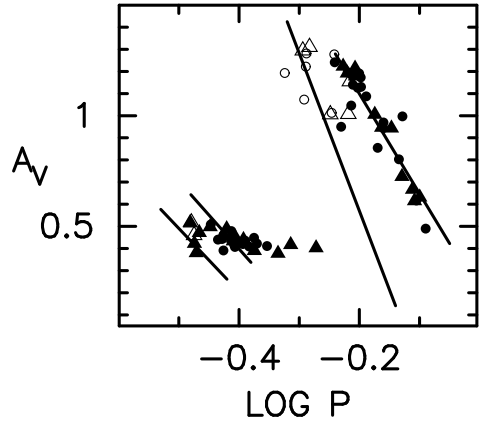}{5.0 truein}{0}{80}{80}{-200}{-10}
\caption{
Period - $V$ amplitude plot
for selected fundamental mode and first overtone RR Lyrae variables in 
$\omega$ Cen. 
Among the fundamental mode variables, only the ones with `normal' light
curves are plotted.
For the overtone pulsators, only those with an error less than $0.2$ in
$\phi_{31}$ are included.
The circles represent the stars in field 5139B: solid
circles for stars with $<V> \le 14.65$ and open circles for the
fainter stars. The solid and open triangles represent stars brighter
and fainter than  $14.65$
in other fields. 
Also included are lines delineating the P-A relations for RRc and RRab stars
in OoI and OoII clusters.
The two lines in the range of $\log P$ between $-0.5$ and $-0.4$ follow
the P-A relation for RRc stars in the OoI cluster M3 and the OoII cluster M68.
The two lines in the range of $\log P$ between $-0.3$ and $-0.1$ represent
least squares fits to the 
P-A relation for OoI RRab stars in M3 and the OoII RRab stars in $\omega$ Cen,
i.e. those with $<V> \le 14.65$.
}
\end {figure}

\clearpage

%

\begin{deluxetable}{ccccccc}
\tablecaption{ The RR Lyrae Variables in $\omega$ Centauri (OGLE data) \label{Table 1} }
\tablewidth{0pt}
\tablehead{
\colhead{Ogle} & \colhead{Sawyer} & \colhead{Field} & \colhead{Period}
& \colhead{$A_V$} & \colhead{$<V>$} & \colhead{Pulsation}\\
\colhead{\#} & \colhead{Hogg \#} & \colhead{} & \colhead{(days)} & \colhead{} 
& \colhead{} &\colhead{mode}
}
\startdata
 71  & 185 & A &    0.3330  &  0.231 & 14.580  & 2nd \nl
 73  & 68 & A &    0.5346 &   0.403 & 14.269 & 1st \nl 
 74  & 54 & A &    0.7729 &   0.667 & 14.460 & F(n) \nl 
 75  & 130 & A &    0.4932 &   0.754 & 14.766 & F(p) \nl 
 76  & 66 & B &    0.4075 &   0.435 & 14.543 & 1st \nl 
 77  & 67 & A &   0.5644 &   1.006 & 14.756 & F(n) \nl 
 78  & 35 & B &    0.3868 &   0.479 & 14.552 & 1st \nl 
 79  &  76 & A &    0.3380 &   0.381 & 14.553 & 1st \nl 
 80  & 77 & B &    0.4264 &   0.422 & 14.535 & 1st \nl 
 81  && A &   0.2496 &   0.089 & 14.615 & 3rd? \nl 
 82  & 32 & A &    0.6204 &   1.214 & 14.603 & F(n) \nl 
 83  & 83 & A &    0.3566 &   0.498 & 14.617 & 1st \nl 
 84  &  75 & A &    0.4221 &   0.390 & 14.509 & 1st \nl 
 85  & 74 & A &    0.5032 &   1.294 & 14.703 & F(n) \nl 
 86  & 36 & A &    0.37983 &  0.489 & 14.558 & 1st \nl 
 87  & 7 & A &    0.7130  &  0.944 & 14.633 & F(n) \nl 
 88  & 44 & B &    0.5675 &    1.013 & 14.763 & F(n) \nl 
 89  & 115 & B &    0.6304 &   1.191 & 14.589 & F(n) \nl 
 90  & 34 & B &    0.7340 &   0.803 & 14.499 & F(n) \nl 
 91  & 128 & B &   0.8350 &   0.610 & 14.332 & F(p) \nl 
 92  & 30 & B &    0.4040 &   0.420 & 14.464 & 1st \nl 
 93  & 59 & B &    0.5184 &   0.868 & 14.739 & F(p) \nl 
 94  & 117 & B &    0.4216 &   0.448 & 14.470 & 1st \nl 
 97  & 21 & B &    0.3808 &   0.484 & 14.385 & 1st \nl 
 98  & 10 & B &    0.3751 &   0.391 & 14.466 & 1st \nl 
 99  & 4 & B &    0.6273  &  1.126 & 14.483 & F(n) \nl 
100 & 58 & B &   0.3699  &  0.288 & 14.441 & 1st \nl
101 & 5 & B &   0.5154  &  1.281 & 14.747 & F(n) \nl 
102 & 122 & B &   0.6349 &   1.130 & 14.570 & F(n) \nl 
104 & 120 & B &   0.5486 &   0.820 & 14.752 & F(p) \nl 
105 & 121 & B &   0.3042 &   0.284 & 14.581 & 2nd \nl
106 & 119 & B &   0.3059 &   0.292 & 14.646 & 2nd \nl
107 & 40 & B &   0.6341  &  1.173 & 14.599 & F(n) \nl 
108 && B &   0.3214  &  0.434 & 14.753 & 1st \nl
109 & 118 & B &   0.6116 &   1.046 & 14.471 & F(n) \nl 
110 & 131 & B &   0.3920 &   0.407 & 14.484 & 1st \nl 
111 && B &   0.4229  &  0.354 & 14.473 & 1st \nl 
112 & 144 & B &   0.8352 &   0.486 & 14.420 & F(p) \nl 
113 & 25 & B &   0.5885  &  0.950 & 14.547 & F(n) \nl 
114 & 27 & B &   0.6157  &  0.658 & 14.749 & F(p) \nl 
115 && B &   0 3524  &  0.255 & 14.539  & 1st \nl 
117 & 62 & B &   0.6198  &  1.184 & 14.523 & F(n) \nl 
118 & 139 & B &   0.6768 &   0.855 &  14.377 & F(n) \nl 
119 & 137 & B &   0.3342 &   0.466 & 14.550  & 1st \nl 
120 & 26 & B &   0.7846 &   0.617 & 14.520 & F(n) \nl 
121 && B &   0.3159  &  0.246 & 14.470 & 2nd \nl
122 && B &   0.8131  &  0.490  & 14.446 & F(n) \nl 
124 & 15 & B &   0.8106  &  0.745 & 14.402 & F(p) \nl 
125 & 12 & B &   0.3868  &  0.453 & 14.534 & 1st \nl 
126 & 11 & B &   0.5643  &  0.737 & 14.539 & F(p) \nl
127 & 116 & B  &  0.7200  &  0.673 & 14.291 & F(p) \nl 
128 & 113 & B &   0.5733 &   1.277 & 14.664 & F(n) \nl 
129 & 41 & B &   0.6630  &  1.013 & 14.584 & F(p) \nl 
130 & 145 & B &   0.3732 &   0.443 & 14.566 & 1st \nl
131 && B &  0.2381 &   0.204 & 14.609  & 3rd? \nl
132 && B &   0.4431 &   0.411 & 14.441 & 1st \nl 
133 & 109 & B &   0.7439 &   0.997 & 14.487 & F(n) \nl 
134 && B &   0.3115  &  0.185 & 14.626 & 2nd \nl
135 & 158 & B &   0.3673 &   0.440 & 14.589 & 1st \nl
136 & 111 & B &   0.7631 &   0.595 & 14.476 & F(p) \nl 
137 & 99 & B &   0.7660  &  1.148 & 14.389 & F(p) \nl 
138 & 157 & B &   0.4066 &   0.435 & 14.571 & 1st \nl 
139 && B &   0.3672  &  0.384 & 14.606 & 1st \nl 
140 & 153 & B &   0.3863 &   0.438 & 14.560 & 1st \nl 
141 & 90 & B &   0.6034  &  1.196 & 14.575 & F(n) \nl 
142 & 166 & B &   0.3414  &  0.117 & 14.545 & 2nd \nl 
143 & 89 & B &   0.3748  &  0.458 & 14.579 & 1st \nl 
144 & 87 & B &   0.3965  &  0.467 & 14.599 & 1st \nl 
145 & 155 & B &   0.4139 &   0.409 & 14.476 & 1st \nl 
146 & 23 & B &   0.5109  &  1.073 & 14.841 & F(n) \nl 
147 & 51 & B &   0.5741  &  1.241 & 14.567 & F(n) \nl 
148 && B &   0.3111  &  0.222 & 14.598 & 2nd \nl  
149 && B &  0.378175 &  0.346 & 14.512 & 1st \nl
150 & 112 & B &   0.4743 &   1.193 & 14.655 & F(n) \nl 
151 & 70 & B &   0.3906  &  0.418 & 14.523 & 1st \nl 
152 && B &   0.3078  &  0.208 & 14.550 & 2nd \nl
153 & 102 & B &   0.6914 &   0.944 & 14.571 & F(n) \nl 
154 & 107 & B &   0.5141 &   1.222 & 14.853 & F(n) \nl 
155 && B &   0.3517  &  0.158 & 14.595  & 2nd \nl 
157 &  71 & B &   0.3574  &  0.508 & 14.591 & 1st \nl 
158 & 86 & B &   0.6478  &  1.087 & 14.583 & F(n) \nl 
159 & 97 & B &   0.6919  &  0.969 & 14.559 & F(n) \nl 
160 & 98 & B &  0.2806 &   0.493 & 14.857 & 1st \nl
162 && B &   0.2817 &   0.153 & 14.718 & 2nd \nl  
163 & 20 & B &   0.6156  &  1.140 & 14.608 & F(n) \nl 
165 & 49 & C &   0.60463 &  1.007 & 14.726 & F(n) \nl
166 & 125 & C &   0.59287 &  1.223 & 14.650 & F(n) \nl 
167 && C &   0.2850  &  0.073 & 14.634 & 2nd \nl 
168 & 124 & C &   0.33185 &  0.517 & 14.641 & 1st \nl
169 & 123 & C &   0.4743 &   0.397 & 14.471 & 1st \nl 
170 & 38 & C &   0.77904 &  0.615 & 14.525 & F(n) \nl 
174 & 84 & D & 0.5799   &   0.638 & 14.291 & F(n) \nl 
175 & 127 & D &   0.3053 &   0.321 & 14.640 & 2nd \nl
176 & 85 & D &   0.74275 &  0.725 & 14.500 & F(n) \nl 
177 && D &   0.3290  &  0.188 & 14.580 & 2nd \nl
178 & 95 & D &   0.4050  &  0.440 & 14.576 & 1st \nl 
179 & 45 & D &   0.58914 &  1.087 & 14.625 & F(p) \nl 
180 & 46 & D &   0.68696 &  0.947 & 14.553 & F(n) \nl 
181 & 168 & D &   0.3213 &   0.445 & 15.181 & 1st \nl
182 & 33 & D &   0.60234 &  1.192 & 14.603 & F(n) \nl 
183 & 9 & BC & 0.5234  &  0.739 & 14.848 & F(p) \nl 
184 & 3 & D &   0.84126  & 0.768 & 14.436 & F(p) \nl 
185 &  47 & D &   0.48505 &   0.417 & 14.362 & 1st \nl 
186 && D &   0.2539 &   0.244 & 14.801 & 2nd \nl 
187 & 50 & D &   0.38616 &  0.450 & 14.672 & 1st \nl 
188 & 13 & D &   0.66905 &  1.006 & 14.544 & F(n) \nl 
189 & 24 & E &   0.4623  &  0.378 & 14.484 & 1st \nl 
190 & 22 & E &   0.3960  &  0.425 & 14.567 & 1st \nl 
191 & 184 & E &   0.3034  &  0.212 & 14.651 & 2nd \nl 
192 & 19 & E &   0.29955 &   0.468 & 14.871 & 1st \nl 
193 &  82 & E &   0.3358 &   0.423 & 14.595 & 1st \nl 
194 & 101 & E & 0.3410 & 0.391 & 14.865 & 1st \nl
195 & 81 & E &   0.3894  &  0.434 & 14.578 & 1st \nl 
196 && B &   0.3191  &  0.227 & 14.597 & 2nd \nl
197 & 39 & E &   0.3934  &  0.446 & 14.579 & 1st \nl 
198 & 105 & E &   0.3353 &   0.460 & 14.743 & 1st \nl 
199 & 8 & E &   0.5213  &  1.310 & 14.752 & F(n) \nl 
200 & 18 & E &   0.6217  &  1.179 & 14.601 & F(n) \nl 
201 & 104 & E &   0.8666 &   0.352 & 14.494 & F(p) \nl 
202 && E &   0.2954  &  0.106 & 14.640 & 2nd \nl 
203 && E &   0.3080  &  0.250 & 14.638 & 2nd \nl
204 & 79 & E &   0.6083  &  1.155 & 14.654 & F(n) \nl 
205 & 16 & F &   0.3302  &  0.517 & 14.566 & 1st \nl
206 & 57 & F &   0.7946 &   0.632 & 14.485 & F(n) \nl 
207 & 126 & F &   0.3420  &  0.472 & 14.616 & 1st \nl 
209 &&  BC &  0.3342 &   0.150 & 14.602 & 2nd \nl 
210 & 88 & BC &  0.6904  &  0.679 & 14.258 & F(p) \nl 
211 & 108 & BC &  0.5945 &   1.147 &  14.664 & F(p) \nl 
\enddata
\end{deluxetable}

\clearpage
\begin{deluxetable}{lcccccccc}
\tablecaption{Fourier Parameters for the Overtone Pulsators in $\omega$
Centauri \label{Table 2} }
\tablewidth{0pt}
\tablehead{
\colhead{Ogle \#} & \colhead{$\log P$} & \colhead{$\phi_{21}$($\sigma$)} & 
\colhead{$\phi_{31}$($\sigma$)} & \colhead{$A_4$}
& \colhead{$M_V$}& \colhead{$\log L/{\lsun}$}  & \colhead{$M/\msun$} 
&\colhead{[Fe/H]$_{hk}$}\\  
}
\startdata
131 & -0.623 & 5.03 (0.37) & 4.45 (0.50) & 0.00626 & & & & \nl
81  & -0.603 & 3.00 (0.34) & 6.99 (1.11) & 0.00221 & & & & \nl
186 & -0.595 & 4.09 (0.32) & 6.74 (3.75) & 0.00295 & & & & \nl
160 & -0.552 & 4.21 (0.27) & 4.49 (0.98) & 0.01091 & & & & -1.05 \nl
162 & -0.550 & 5.99 (0.52) & 7.50 (2.79) & 0.00180 & & & & \nl
167 & -0.545 & 4.93 (0.72) & 4.94 (1.07) & 0.00136 & & & & \nl
202 & -0.530 & 3.96 (0.52) & 8.21 (0.77) & 0.00396 & & & & \nl
192 & -0.524 & 4.66 (0.03) & 3.34 (0.12) & 0.00936 & 0.80 & 1.67 & 0.56 &
   -1.22 \nl
191 & -0.518 & 4.68 (0.10) & 3.38 (0.56) & 0.00123 & 0.83 & & & \nl
105 & -0.517 & 4.81 (0.11) & 3.81 (0.51) & 0.00528 & 0.80 & & & -1.46  \nl
175 & -0.515 & 4.87 (0.23) & 4.60 (0.35) & 0.00505 & & &  & -1.59 \nl
106 & -0.514 & 5.02 (0.18) & 3.53 (0.28) & 0.00483 & 0.79 & & & -1.61 \nl
152 & -0.512 & 4.47 (0.24) & 2.75 (0.71) & 0.00075 & & & & \nl
203 & -0.511 & 4.80 (0.24) & 7.66 (0.74) & 0.00313 & & & & \nl
148 & -0.507 & 4.30 (0.33) & 3.54 (0.40) & 0.00094 & & & & \nl
134 & -0.507 & 4.85 (0.23) & 4.24 (2.19) & 0.00475 & & & & \nl
121 & -0.501 & 5.57 (0.56) & 2.71 (0.51) & 0.00373 & & & & \nl
196 & -0.496 & 4.78 (0.18) & 2.67 (0.53) & 0.00067 & 0.81 & & & \nl
181 & -0.493 & 4.81 (0.04) & 3.43 (0.05) & 0.01112 & 0.76 & 1.70 & 0.57 & 
\nl
108 & -0.493 & 4.67 (0.16) & 3.96 (0.24) & 0.00859 & 0.78 & & & \nl
177 & -0.483 & 4.62 (0.32) & 2.33 (0.63) & 0.00377 & & & & \nl
205* & -0.481 & 4.97 (0.10) & 3.38 (0.15) & 0.02481 & 0.68 & 1.71 & 0.59 &
   -1.29 \nl
168* & -0.479 & 4.82 (0.04) & 2.99 (0.07) & 0.02084 & 0.71 & 1.74 & 0.65 &
   -1.33 \nl
209 & -0.476 & 4.93 (0.24) & 7.08 (1.00) & 0.00797 & & & & \nl
198* & -0.475 & 5.13 (0.09) & 4.00 (0.10) & 0.00921 & 0.74 & 1.68 & 0.51 &
   -1.24 \nl
193* & -0.474 & 4.92 (0.04) & 3.24 (0.08) & 0.01198 & 0.74 & 1.73 & 0.61 &
   -1.56 \nl
79*  & -0.471 & 5.02 (0.07) & 3.38 (0.14) & 0.00778 & 0.75 & 1.72 & 0.59 &
   -1.45 \nl
194 & -0.467 & 5.37 (0.52) & 3.02 (0.75) & 0.02575 & & & & -1.88 \nl
142 & -0.467 & 5.19 (0.38) & 4.42 (0.79) & 0.00215 & & & & \nl
207* & -0.466 & 4.98 (0.05) & 3.41 (0.08) & 0.01446 & 0.72 & 1.73 & 0.59 &
   -1.31 \nl
155 & -0.454 & 5.84 (0.54) & 4.76 (0.35) & 0.00399 & & & & \nl
115 & -0.453 & 4.91 (0.26) & 2.07 (0.63) & 0.00536 & & & & \nl
83**  & -0.448 & 4.92 (0.03) & 3.89 (0.04) & 0.01343 & 0.71 & 1.72 & 0.54 &
   -1.30 \nl
157** & -0.447 & 4.78 (0.11) & 3.35 (0.17) & 0.01371 & 0.72 & 1.75 & 0.62 
& \nl
139 & -0.435 & 5.64 (0.21) & 4.11 (0.21) & 0.00932 & & & & \nl
135** & -0.435 & 5.45 (0.14) & 4.44 (0.19) & 0.00441 & 0.72 & 1.70 & 0.47 &
   -1.25 \nl
100 & -0.432 & 5.74 (0.22) & 5.00 (1.24) & 0.00305 & & & & -1.37 \nl
130** & -0.428 & 5.15 (0.08) & 3.44 (0.15) & 0.00817 & 0.71 & 1.77 & 0.62 &
   -1.58 \nl
143** & -0.426 & 5.32 (0.08) & 4.07 (0.12) & 0.00693 & 0.71 & 1.73 & 0.53 &
    -1.37 \nl
98**  & -0.426 & 5.39 (0.04) & 3.79 (0.07) & 0.00896 & 0.69 & 1.75 & 0.56 &
    -1.66 \nl
149 & -0.422 & 5.37 (0.54) & 3.77 (0.52) & 0.01088 & & & & \nl
86**  & -0.420 & 4.92 (0.07) & 4.18 (0.08) & 0.01154 & 0.70 & 1.73 & 0.52 &
    -1.49 \nl
97  & -0.419 & 2.48 (0.21) & 5.22 (0.09) & 0.01279 & 0.80 & 1.67 & 0.40 &
    -0.90 \nl
187** & -0.413 & 5.43 (0.07) & 4.52 (0.10) & 0.01227 & 0.67 & 1.72 & 0.48 &
    -1.59 \nl
140** & -0.413 & 5.24 (0.08) & 3.79 (0.11) & 0.00804 & 0.69 & 1.76 & 0.57 &
    -1.38 \nl
125** & -0.413 & 4.89 (0.19) & 4.30 (0.17) & 0.01157 & 0.69 & 1.73 & 0.50 &
    -1.53 \nl
78**  & -0.413 & 5.16 (0.13) & 4.05 (0.15) & 0.01419 & 0.67 & 1.75 & 0.54 &
    -1.56 \nl
195** & -0.410 & 5.06 (0.08) & 4.33 (0.10) & 0.00759 & 0.70 & 1.73 & 0.50 &
    -1.72 \nl
151** & -0.408 & 5.46 (0.10) & 4.21 (0.07) & 0.00975 & 0.67& 1.74 & 0.52 &
    -1.94 \nl
110** & -0.407 & 5.24 (0.09) & 4.23 (0.10) & 0.00881 & 0.68 & 1.74 & 0.52 &
    -1.56 \nl
197** & -0.405 & 5.16 (0.08) & 4.41 (0.09) & 0.00957 & 0.68 & 1.73 & 0.50 &
    -1.96 \nl
190** & -0.402 & 5.38 (0.08) & 4.38 (0.07) & 0.01098 & 0.66 & 1.74 & 0.50 &
    -1.93 \nl
144 & -0.402 & 5.10 (0.18) & 3.96 (0.21) & 0.00899 & 0.69 & & & -1.44 \nl
92**  & -0.394 & 5.32 (0.18) & 4.50 (0.15) & 0.00945 & 0.67 & 1.74 & 0.49 &
    -1.75 \nl
178** & -0.393 & 5.39 (0.11) & 4.15 (0.13) & 0.01237 & 0.65 & 1.76 & 0.54 &
    -1.84 \nl
138 & -0.391 & 5.23 (0.53) & 4.32 (0.40) & 0.01146 &  & & & -1.49\nl
76**  & -0.390 & 5.37 (0.10) & 4.16 (0.10) & 0.00957 & 0.66 & 1.76 & 0.54 &
    -1.68 \nl
145** & -0.383 & 5.29 (0.11) & 4.26 (0.13) & 0.00928 & 0.66 & 1.76 & 0.53 &
    -1.46 \nl
94**  & -0.375 & 5.49 (0.12) & 4.48 (0.10) & 0.01571 & 0.61 & 1.76 & 0.50 &
     -1.68 \nl
84**  & -0.375 & 5.88 (0.12) & 4.71 (0.13) & 0.00713 & 0.63 & 1.75 & 0.48 &
     -1.49 \nl
111 & -0.374 & 5.42 (0.48) & 5.12 (0.68) & 0.00454 &  & & & \nl
80**  & 0.370 & 5.60 (0.12) & 4.20 (0.17) & 0.01143 & 0.62 & 1.78 & 0.54 &
     -1.81 \nl
132 & -0.354 & 7.50 (0.37) & 5.13 (0.13) & 0.01449 & & 1.745 & 0.44 & \nl
189 & -0.335 & 6.33 (0.09) & 4.49 (0.08) & 0.00508 & 0.59 & 1.80 & 0.53 & 
\nl
169 & -0.324 & 8.35 (0.14) & 5.48 (0.23) & 0.00648 & 0.48 & & & -1.64 \nl
185 & -0.314 & 8.10 (0.12) & 5.40 (0.12) & 0.00692 & 0.48 & 1.77 & 0.43 &
     -1.58 \nl
73  & -0.272 & 8.38 (0.09) & 5.21 (0.08) & 0.00573 & 0.42 & 1.83 & 0.47 &
     -1.60 \nl
\enddata
\end{deluxetable}

\clearpage
\begin{deluxetable}{cccccc}
\tablecaption{Fourier Parameters for RRab Stars in $\omega$ Centauri
\label{Table 3} }
\tablewidth{0pt}
\tablehead{
\colhead{Ogle \#} & \colhead{$\log P$} &  
 \colhead{$A_1$} & \colhead{$\phi_{31}$($\sigma$)} & 
 \colhead{$M_V$} & [Fe/H]$_{hk}$ }
\startdata
 150  & -0.324 & 0.392 & 1.51 & 0.85  & -1.81 \nl
 85   & -0.298 & 0.465 & 1.55 & 0.78  & -1.83 \nl
 146  & -0.292 & 0.364 & 1.85 & 0.85  & -1.08 \nl
 154  & -0.289 & 0.399 & 1.81 & 0.82  & -1.36 \nl
 101  & -0.288 & 0.441 & 1.79 & 0.80  & -1.35 \nl
 199  & -0.283 & 0.452 & 1.58 & 0.76  & -1.91 \nl
 77   & -0.248 & 0.358 & 1.96 & 0.79  & -1.10 \nl
 88   & -0.246 & 0.343 & 2.05 & 0.80  & -1.40 \nl
 128  & -0.242 & 0.437 & 1.77 & 0.72  & -1.65 \nl
 147  & -0.241 & 0.426 & 1.77 & 0.72  & -1.64 \nl
 174  & -0.237 & 0.239 & 2.32 & 0.86  & -1.47 \nl
 113  & -0.230 & 0.396 & 1.75 & 0.71  & -1.57 \nl 
 166  & -0.227 & 0.414 & 1.88 & 0.71  & -1.67 \nl
 182  & -0.220 & 0.398 & 1.90 & 0.71  & -2.09 \nl
 141  & -0.219 & 0.398 & 1.90 & 0.71  & -2.21 \nl
 165  & -0.219 & 0.344 & 1.83 & 0.73  & -1.98 \nl
 204  & -0.216 & 0.391 & 1.98 & 0.71  & -1.39 \nl
 109  & -0.214 & 0.349 & 1.87 & 0.72  & -1.62 \nl
 163  & -0.211 & 0.391 & 1.97 & 0.70  & \nl
 117  & -0.208 & 0.399 & 1.91 & 0.69  & -1.62 \nl
 82   & -0.207 & 0.407 & 1.93 & 0.68  & -1.53 \nl
 200  & -0.206 & 0.396 & 1.95 & 0.69  & -1.78 \nl
 99   & -0.203 & 0.381 & 1.98 & 0.69  & -1.74 \nl
 89   & -0.200 & 0.385 & 1.93 & 0.68  & -1.87 \nl
 107  & -0.198 & 0.379 & 2.00 & 0.68  & -1.60 \nl
 102  & -0.197 & 0.376 & 1.98 & 0.68  & -2.02 \nl
 158  & -0.189 & 0.360 & 2.06 & 0.68  & -1.81 \nl
 188  & -0.175 & 0.365 & 2.31 & 0.67  & -1.91 \nl
 118  & -0.170 & 0.289 & 2.12 & 0.68  & -1.46 \nl
 180  & -0.163 & 0.323 & 2.23 & 0.66  & -1.88 \nl
 153  & -0.160 & 0.328 & 2.18 & 0.65  & -1.84 \nl
 159  & -0.160 & 0.330 & 2.21 & 0.65  & -1.56 \nl
 87   & -0.147 & 0.328 & 2.26 & 0.63  & -1.46 \nl
 90   & -0.134 & 0.287 & 2.36 & 0.63  & -1.71 \nl
 176  & -0.129 & 0.270 & 2.53 & 0.64  & -1.87 \nl
 133  & -0.129 & 0.355 & 2.45 & 0.59  & -1.51 \nl
 74   & -0.112 & 0.250 & 2.68 & 0.62  & -1.66 \nl
 170  & -0.108 & 0.230 & 2.72 & 0.63  & -1.75 \nl
 120  & -0.105 & 0.237 & 2.63 & 0.61  & -1.68 \nl
 206  & -0.100 & 0.228 & 2.66 & 0.60  & -1.89 \nl
 122  & -0.090 & 0.192 & 2.67 & 0.59  & \nl

\enddata
\end{deluxetable}

\clearpage
\begin{deluxetable}{lccccccc}
\tablecaption{Derived Parameters for RRc Stars of 
Different Oosterhoff Types \label{Table 4} }
\tablewidth{0pt}
\tablehead{
\colhead{Cluster} & \colhead{[Fe/H]} &  \colhead{No. of } &
\colhead{mean P} & \colhead{Oosterhoff } &  
 \colhead{mean} & \colhead{mean} & 
 \colhead{mean} \\ 
&& \colhead{Stars} 
  & \colhead{(days)} & \colhead{type} &  
 $M/\msun$ & \colhead{$\log L/\lsun$} & $T_{eff}$  }
\startdata
M107 (N6171) & -0.99 & 6 & 0.292 &  I & $0.54\pm 0.01$ & $1.65\pm 0.01$ & 
$7448\pm19$ \nl
M5   & -1.40 & 14 & 0.324 & I & $0.54\pm 0.02$ & $1.69\pm 0.01$ & $7353\pm 19$
\nl
M3   & -1.66 & 5 & 0.325  &  I & $0.59\pm 0.03$ & $1.71\pm 0.01$ & $7315\pm 7$
\nl
$\omega$ Cen & (-1.36) & 6  & 0.336 &  I  & $0.59\pm 0.02$ & $1.72\pm 0.01$ & 
$7287\pm 12$ \nl
$\omega$ Cen & (-1.60) & 23 & 0.391 &  II & $0.53\pm 0.01$ & $1.74\pm 0.005$ & 
$7199\pm 8$ \nl
M55 & -1.82 & 5 & 0.391 &    II & $0.53\pm 0.01$ & $1.75\pm 0.01$ & $7193\pm 12$
\nl
M68 & -2.09 & 16 & 0.369  & II & $0.71\pm 0.01$ & $1.79\pm 0.01$ & $7145\pm 18$
\nl
M15 & -2.15 & 6 & 0.367 &   II & $0.73\pm 0.02$ & $1.80\pm 0.01$ & $7136\pm 31$
\nl
\enddata
\end{deluxetable}

\clearpage
\begin{deluxetable}{cccccccc}
\footnotesize
\tablecaption{Mass Differences for Pairs of RRab Stars in $\omega$ Centauri
\label{Table 5} }
\tablewidth{0pt}
\tablehead{
\colhead{Stars} 
& \colhead{$\Delta \log  P$} &  \colhead{$\Delta V$} & \colhead{$\Delta A_V$} &
\colhead{$\Delta$[Fe/H]} & \colhead{$\Delta \log M/\msun$} 
& \colhead{$\Delta \log M/\msun$} & \colhead{$\Delta \log M/\msun$} \\
\colhead{OoI, OoII} & \colhead{(II-I)} 
& \colhead{(II-I)}
& \colhead{(II-I)}
& \colhead{(II-I)}
& \colhead{($\Delta T=0.000$)\tablenotemark{\dag}}&  
\colhead{($\Delta T=-0.004$)\tablenotemark{\dag}}&    
\colhead{($\Delta T=-0.005$)\tablenotemark{\dag}}   \\}
\startdata
150, 141 & 0.1046 & -0.080 &  0.003 & -0.60 & -0.107 & -0.089 &  -0.077 \nl
150, 117 & 0.1162 & -0.132 & -0.009 & -0.11 & -0.110 & -0.091 &  -0.080 \nl 
146, 158 & 0.1031 & -0.258 &  0.014 & -0.73 & -0.009 &  +0.003 & +0.015 \nl
154, 147 & 0.0479 & -0.286 &  0.019 & -0.28 & +0.079 &  +0.093 & +0.105 \nl
101, 128 & 0.0462 & -0.083 &  0.004 & -0.30 & -0.022 & -0.005 & +0.008 \nl
88, 133  & 0.1175 & -0.276 & -0.016 & -0.11 & -0.037 & -0.022 & -0.011 \nl
113, 153 & 0.0700 & +0.024 & -0.006 & -0.27 & -0.114 & -0.096 & -0.083 \nl
mean  & 0.0865 & -0.156 && -0.34 & -0.046 & -0.030 & -0.018 \nl
\enddata
\tablenotetext{\dag}{$\Delta T$ refers to $\Delta \log T_e$} 
\end{deluxetable}

\end{document}